\documentclass[twocolumn]{aastex63}

\received{}
\revised{}
\accepted{}

\shorttitle{SL analysis of SMACS0723}
\shortauthors{Golubchik et al.}

\begin{document}

\title{HST strong-lensing model for the first JWST galaxy cluster SMACS J0723.3-7327}

\correspondingauthor{Adi Zitrin}
\email{adizitrin@gmail.com}

\author[0000-0001-9411-3484]{Miriam Golubchik}
\author[0000-0001-6278-032X]{Lukas J. Furtak}
\author[0000-0002-7876-4321]{Ashish K. Meena}
\author[0000-0002-0350-4488]{Adi Zitrin}
\affiliation{Physics Department,
Ben-Gurion University of the Negev, P.O. Box 653,
Be'er-Sheva 84105, Israel}



\begin{abstract}
On 2022 July 8, NASA shared\footnote{\url{https://www.nasa.gov/feature/goddard/2022/nasa-shares-list-of-cosmic-targets-for-webb-telescope-s-first-images}} the list of five public showcase targets which have been observed with the new \textit{James Webb Space Telescope} (JWST), and whose data are expected to be released to the public around Tuesday, July 12. One of these targets is the galaxy cluster SMACS~J0723.3-7327 which acts as a gravitational lens and was recently imaged with the \textit{Hubble Space Telescope} in the framework of the \textit{Reionization Lensing Cluster Survey} program (RELICS). To facilitate studies by the community with the upcoming JWST data, we publish here a lens model for SMACS0723 -- including mass-density and magnification maps. We identify in the HST imaging five multiple-image families for three of which membership and redshift are secured by public spectroscopic data. For the remaining two systems we rely on robust photometric redshift estimates. We use here the \texttt{Light-Traces-Mass} lens modeling method, which complements the parametric models already available on the RELICS website and elsewhere, and thus helps span a representative range of solutions. The new models published here can be accessed via a link given below\footnote{\url{https://www.dropbox.com/sh/svc84qmnlh05lmc/AABFQs_0VXE6Fh8jWbAgh4sza?dl=0}}. It will be interesting to examine by how much and which properties of the mass models change, and improve, when JWST data are incorporated.
\end{abstract}




\section{Introduction}\label{sec:intro}

\begin{figure*}
 \begin{center}
  \includegraphics[width=190mm,height=170mm,trim=4cm 1cm 1cm 1cm,clip]{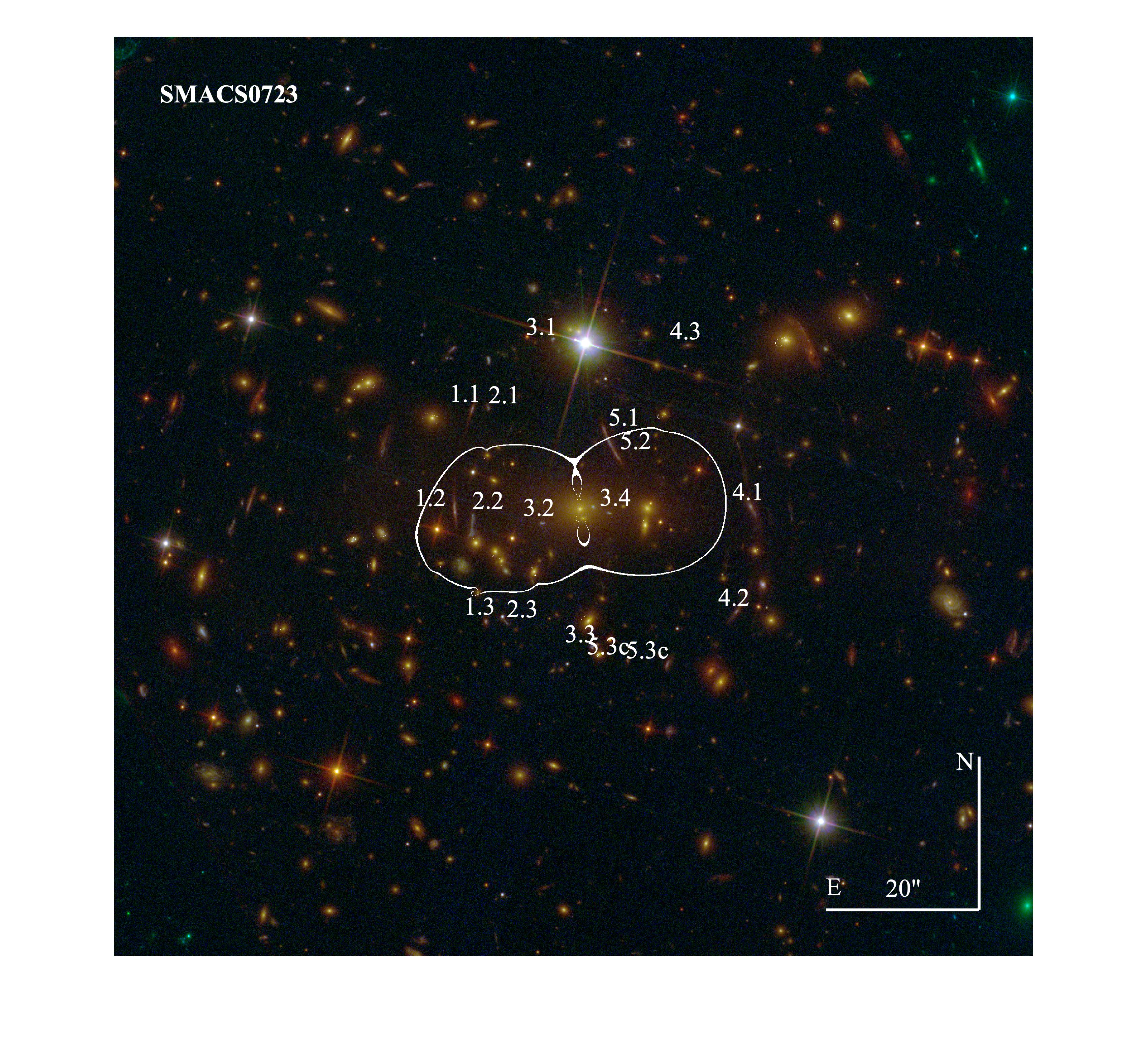}
 \end{center}
\caption{Critical curves and multiple images for SMACS0723 from the \texttt{LTM} model (result of this work). Shown is a color-composite image from the HST data. Multiple-image sets are marked on the image and the critical curves from our model are overlaid in \emph{white} for a source at the redshift of system 1, $z_{\mathrm{spec}}\simeq1.45$). Note also that the ellipticity of the mass distribution and critical curves is limited in the LTM model, resulting in a high external shear implied for this cluster. While we leave a comparison with parametric models for future works, we anticipate that the critical curves would be somewhat more elongated that seen here.}\vspace{0.1cm}
\label{fig:curve}
\end{figure*}

After much anticipation, the first images from the \textit{James Webb Space Telescope} (JWST) are expected to be released starting July 12, 2022. According to the media, these will first include five targets whose names were published last week by NASA, and will be followed by a release of observations taken for Director's Discretionary Early Release programs. One of the first five targets is SMACS~J0723.3-732 (SMACS0723 hereafter), a massive strong-lensing (SL) galaxy cluster which exhibits various gravitationally lensed arcs as seen in recent \emph{Hubble Space Telescope} (HST) imaging taken in the framework of the \textit{Reionization Lensing Cluster Survey} \citep[RELICS;][]{Coe2019RELICS}. Some of the related media on upcoming JWST data mentioned that these images will include the deepest imaging of the Universe taken to date. While a very deep and impressive Fine Guidance Sensor image may have already qualified for the title, suspicion raises that -- thanks to the depth gained by the lensing magnification -- these rumors refer to the announced image of SMACS0723. Either way, a detailed SL model is needed to interpret whatever magnified or strongly lensed features will be seen in these new data.

Over the past 2-3 decades, gravitational lensing has enabled the detection of increasing numbers of faint high-redshift ($z>6$) galaxies \citep[e.g.][]{Franx1997,Kneib2004z7,Bradley2008,Zheng2012NaturZ,Kashikawa2011,Atek2015HFFhalf} which currently represent the best candidates for the sources that reionized the Universe \citep{Atek2015HFFhalf}. The SL magnification allows to probe several magnitudes deeper than blank fields, i.e. down to rest-frame UV luminosities $M_{\mathrm{UV}}\lesssim-13$ magnitudes \citep{bouwens17a,livermore17,atek18,ishigaki18} and stellar masses $M_{\star}\gtrsim10^6\,\mathrm{M}_{\odot}$ \citep{bhatawdekar19,kikuchihara20,furtak21,strait21}. Several high-redshift galaxies have already been detected in SMACS0723 with the HST data from the RELICS program \citep[][]{Salmon2020HighzRelics,strait21}. With the redder wavelength range and greater sensitivity of the JWST, we can expect many more galaxies at even higher redshifts to be detected in the coming months. In order to fully characterize them and study their physics, we will need accurate SL magnification models.

At least two lensing models using the \texttt{glafic} \citep{Oguri2010_25clusters} and \texttt{Lenstool} \citep{Jullo2007Lenstool} codes (see also, e.g. \citealt{FoxMahlerSharon2022}) have already been generated for SMACS0723 and are publicly available on the RELICS website\footnote{\url{https://archive.stsci.edu/prepds/relics/}}. In addition, another \texttt{Lenstool} model was recently constructed (J. Richard, private communication) using similar constraints to those used here, including the three spectroscopic redshifts from MUSE, and is publicly available online as well\footnote{\url{https://cral-perso.univ-lyon1.fr/labo/perso/johan.richard/ALCS_models/SMACS0723/}}. These models are considered \emph{parametric}, in the sense that the cluster galaxies and dark matter (DM) components are assumed to follow known profile shapes which can be modeled using analytic or parametrized formulae. Here, we present a new model with the \texttt{Light-Traces-Mass} approach \citep[\texttt{LTM};][]{Broadhurst2005a,Zitrin2009_cl0024,Zitrin2014CLASH25}. On account of being different in nature than parametric techniques, the \texttt{LTM} method allows to probe a different range of solutions. This will be very important for high-redshift studies in this cluster, for which a representative range of possible magnifications for background galaxies is needed. In addition, given the much deeper JWST data and their longer wavelength coverage, we expect that more lensed galaxies will be uncovered soon using these data, and so the model could be refined further and compared to the existing pre-JWST version.

The paper is organized as follows: In \S \ref{s:data} we detail previous observations of the cluster and their use in the lens modeling, which is described in \S \ref{s:code}. In \S \ref{s:results} we present and discuss the results. The work is concluded in \S \ref{s:summary}. Throughout this work we use a $\Lambda$CDM cosmology with $\Omega_{M}=0.3$, $\Omega_{\Lambda}=0.7$, and $H_{0}=70$ km/s/Mpc. Unless otherwise stated, we generally use AB magnitudes \citep{Oke1983ABandStandards}, and errors correspond to $1\sigma$.

\section{Observations and Data}\label{s:data}
The galaxy cluster SMACS0723 is part of the southern extension of the MACS sample \citep{Ebeling2010FinalMACS,ReppEbeling2018}, and was recently observed in the framework of RELICS.

The RELICS program observed 41 massive galaxy clusters with HST (PI: D. Coe), and the \emph{Spitzer Space Telescopes} (PI: M. Bradac) with the goal of detecting gravitationally lensed arcs, bright high-redshift galaxies \citep{Salmon2020HighzRelics,strait21} and various transients (e.g. supernovae listed in \citealt{Coe2019RELICS} and the first spectacular detection of a lensed star at $z\simeq6.2$ shown in \citealt{welch22}). Each cluster was observed (in two separate epochs) to about AB mag 26.5 in 7 HST bands: F435W, F606W, F814W with the \textit{Advanced Camera for Surveys} (ACS), and F105W, F125W, F140W, and F160W, with the \textit{Wide Field Camera Three} (WFC3). Some clusters also build on previous HST observations \citep{Coe2019RELICS}.

For SMACS0723, additional HST imaging was taken following supernovae detected in its field, as detailed in \citet{Coe2019RELICS}. The RELICS data products for SMACS0723 include reduced and color images, photometric catalogs generated with \texttt{SExtractor} \citep{BertinArnouts1996Sextractor} and photometric redshifts computed with the \texttt{Bayesian Photometric Redshifts} code \citep[BPZ;][]{Benitez2004,Coe2006}. These are publicly available through the RELICS website\footnote{\url{https://relics.stsci.edu/}}. We refer the reader to \citet[][]{Coe2019RELICS} for more details on the HST data reduction and catalog assembly.

In this work, we also make use of data from the \textit{Multi Unit Spectroscopic Explorer} (MUSE; \citealt{Bacon2010MUSE}) on ESO’s \textit{Very Large Telescope} (VLT). SMACS0723 was observed with MUSE for 2910\,s (Program ID 0101.A-0718; PI: A. Edge), and the data were published on 2019, April 16 on the ESO Science Archive\footnote{\url{http://archive.eso.org/scienceportal/home}}. We use these public data to secure multiple image families identified in the HST data and measure their spectroscopic redshifts where possible.

Note that SMACS0723 was also observed with the \textit{Atacama Large Millimeter/submillimeter Array} (ALMA) in ALMA band 6 in the framework of the \textit{ALMA Lensing Cluster Survey} (ALCS; PI: K. Kohno), although these data are not used here.

\section{SL modeling of SMACS0723}\label{s:code}
Our \texttt{LTM} SL model of SMACS0723 takes two components as inputs: The strongly lensed multiple image systems described in section~\ref{sec:multiple-images} as SL constraints and the cluster member galaxies identified in section~\ref{sec:cluster_members} as one cluster mass component. We also give a brief summary of our modeling methods in section~\ref{sec:SL-code}.

\subsection{Multiple images} \label{sec:multiple-images}

\begin{deluxetable*}{lcccc}
\tablecaption{Multiple image systems in SMACS0723}
\label{multTable}
\tablecolumns{5}
\tablehead{
\colhead{ID
} &
\colhead{R.A
} &
\colhead{DEC.
} &
\colhead{$z$
} &
\colhead{$z_{\mathrm{model}}$
} \\  
 &J2000.0&J2000.0&&  
}
\startdata
1.1 & 7:23:21.7192 & -73:27:03.583 & 1.45 & ---  \\
1.2 & 7:23:22.2576 & -73:27:17.164 & 1.45 & --- \\
1.3 & 7:23:21.3073 & -73:27:31.320 &  1.45 & ---\\
\hline
2.1 & 7:23:21.2301 & -73:27:03.719  & 1.38  & ---\\
2.2 & 7:23:21.7216 & -73:27:18.720  & 1.38 & ---\\ 
2.3 & 7:23:20.6700 & -73:27:31.666  & 1.38  & ---\\
\hline
3.1 & 7:23:19.2740& -73:26:54.830  & --- & 1.27 [1.24 -- 1.30]\\
3.2 & 7:23:19.6361& -73:27:18.452  & 1.147  [1.132 -- 1.415]  & "\\
3.3 & 7:23:18.0700& -73:27:34.936  & 1.227  [1.196 -- 2.330]  & "\\
3.4 & 7:23:17.5730& -73:27:17.103  & ---  & "\\
\hline
4.1 & 7:23:13.2336& -73:27:16.371 & 2.123  [2.059 -- 2.191]  & 2.29 [2.15 -- 2.49]\\ 
4.2 & 7:23:13.6690& -73:27:30.162  & 2.239 [2.108 -- 2.276] & "\\
4.3 & 7:23:15.1406& -73:26:55.358 &  2.387 [2.179 -- 2.472]  & " \\
\hline
5.1 & 7:23:17.6757 & -73:27:06.662 & 1.43 &---\\
5.2 & 7:23:17.3237 & -73:27:09.720 &  1.43 &---\\
5.3c & 7:23:15.7790 & -73:27:37.037 &  1.43 &--- \\
5.3c & 7:23:16.9814 & -73:27:36.522 &  1.43 &--- \\
\hline
\enddata
\tablecomments{Multiple images and candidates. \emph{Column~1:} ID; \emph{Column~2 \& 3:} Right Ascension and Declination, in J2000.0; \emph{Column~4:} Redshift. For systems 1, 2 and 3 we quote the spectroscopic redshift (see section~\ref{sec:multiple-images}) without the error. For systems 3 and 4 we quote the best photometric redshift from RELICS (see section~\ref{s:data}), and its 95\% ($2\sigma$) confidence interval; \emph{Column~5:} The redshift of the system as resulting from the SL model. Candidate images whose identification is not secure are marked with ``c". Note that these were therefore not used in the minimization.}
\end{deluxetable*}

\begin{figure}
 \begin{center}
  \includegraphics[width=90mm,height=70mm,trim=1cm 0.5cm 1cm 5cm, clip]{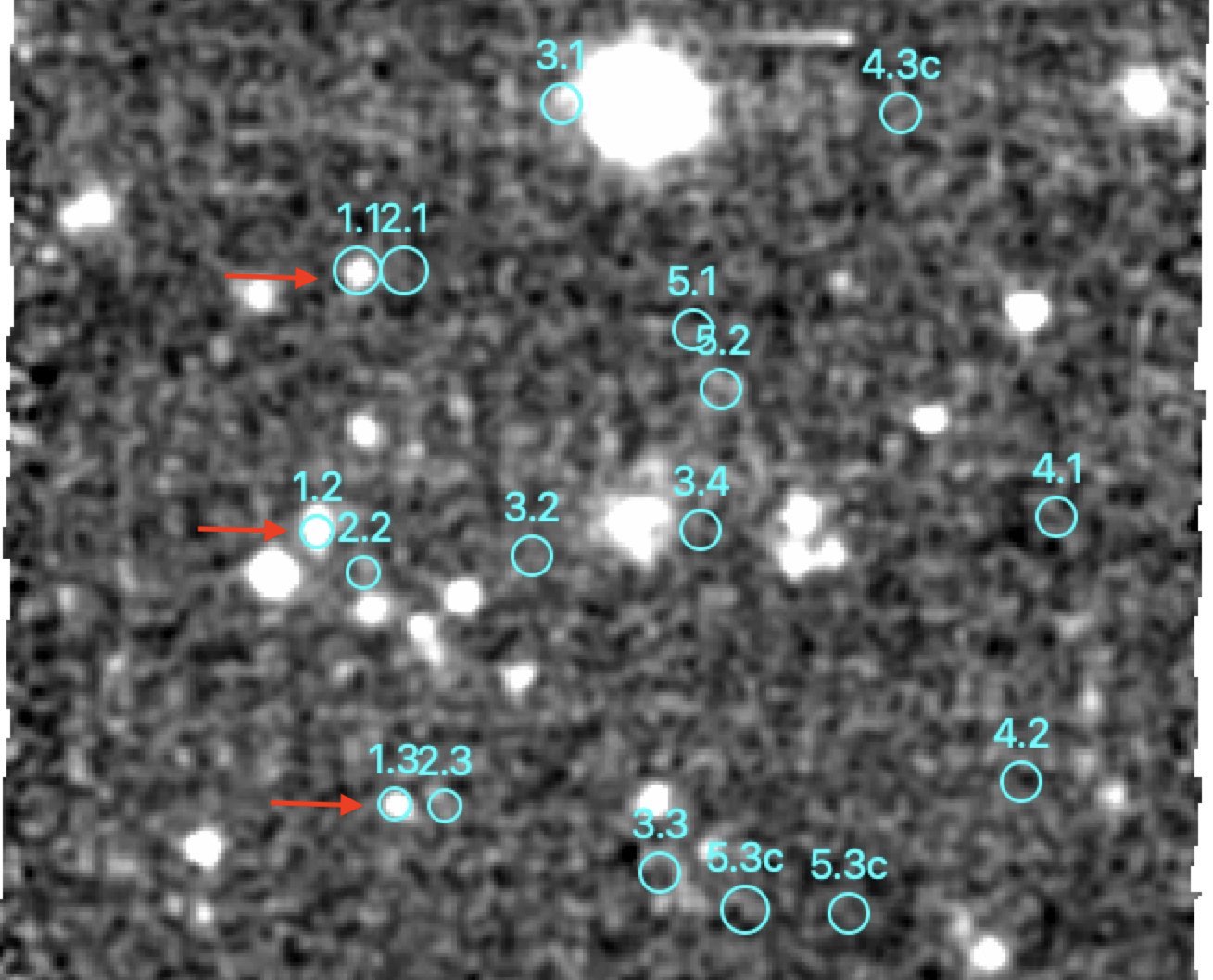}\\
    \includegraphics[width=90mm,height=70mm,trim=1cm 0.5cm 1cm 5cm,clip]{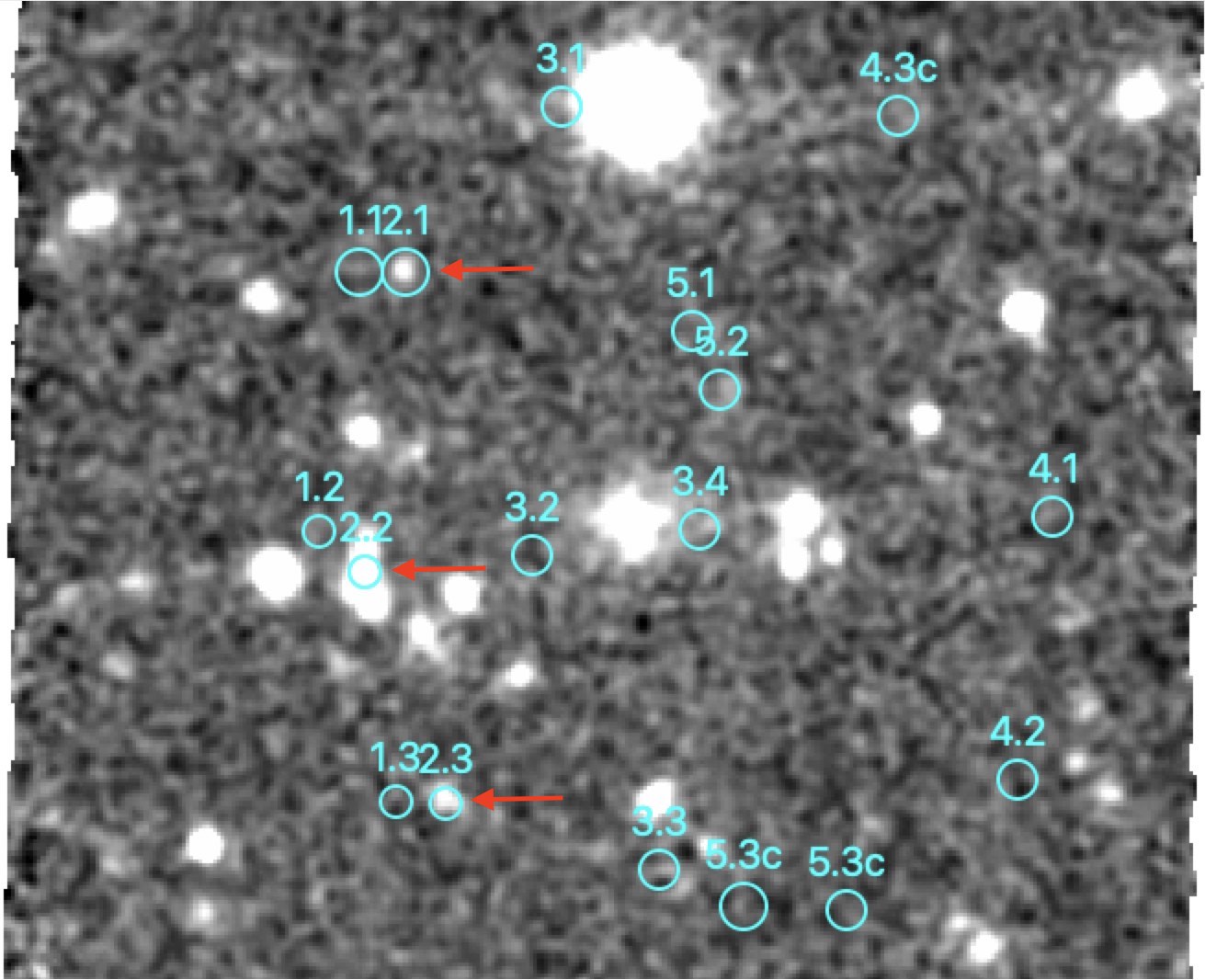}\\
      \includegraphics[width=90mm,height=70mm,trim=1cm 0.5cm 1cm 5cm,clip]{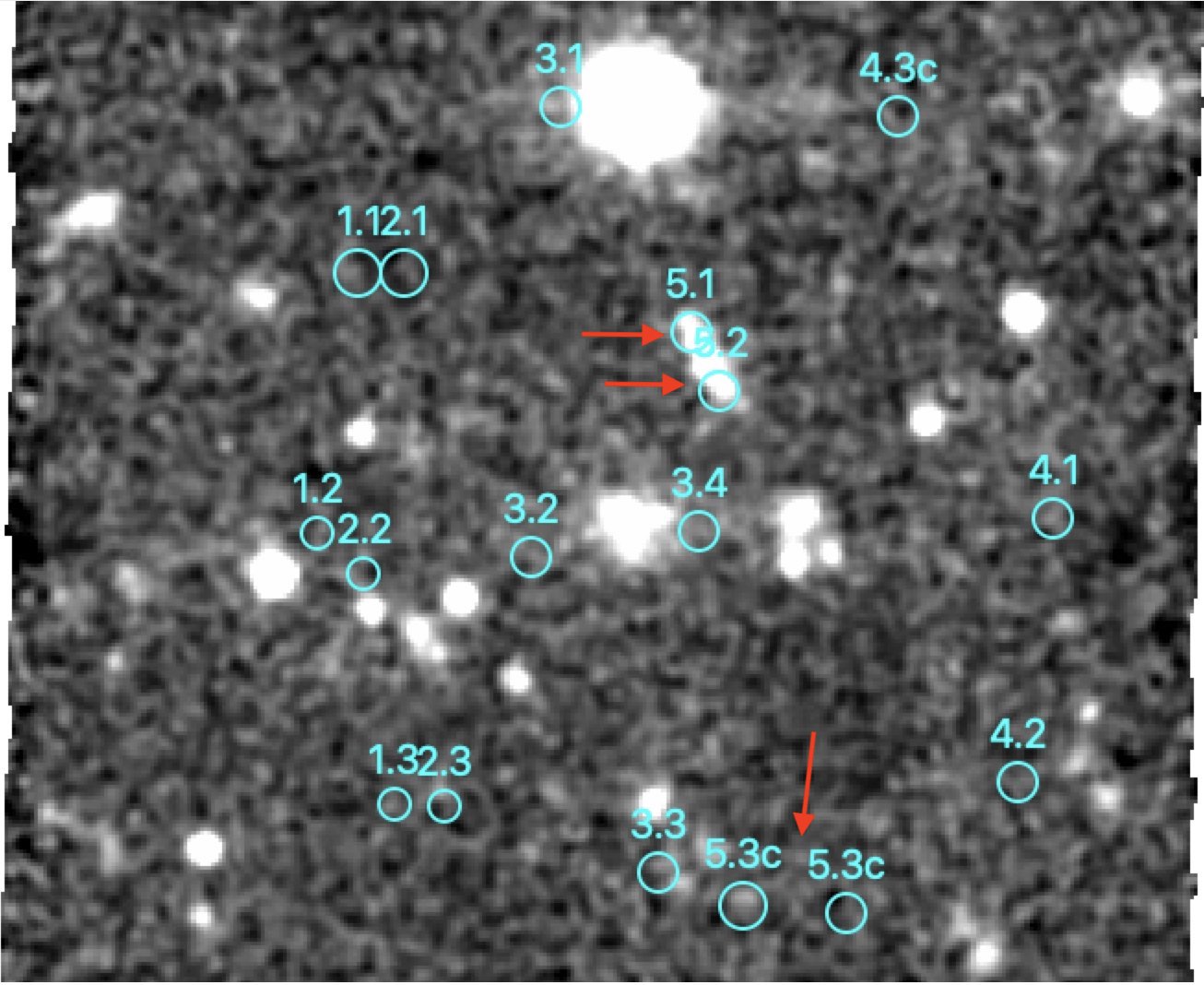}\\
 \end{center}
 \caption{MUSE slices showing the [OII] emission from the three images of system 1 (\emph{top}; $\lambda\simeq9134$\AA); the three images of system 2 (\emph{middle}; $\lambda\simeq8873$\AA); and the two merging images of system 5 (\emph{bottom}; $\lambda\simeq9040$\AA). Also marked are the two most probable candidates for a third counter image in system 5 (5.3c).}
\label{fig:MUSE_images}
\end{figure}

\begin{figure}
    \centering
    \includegraphics[width=\columnwidth, keepaspectratio=true]{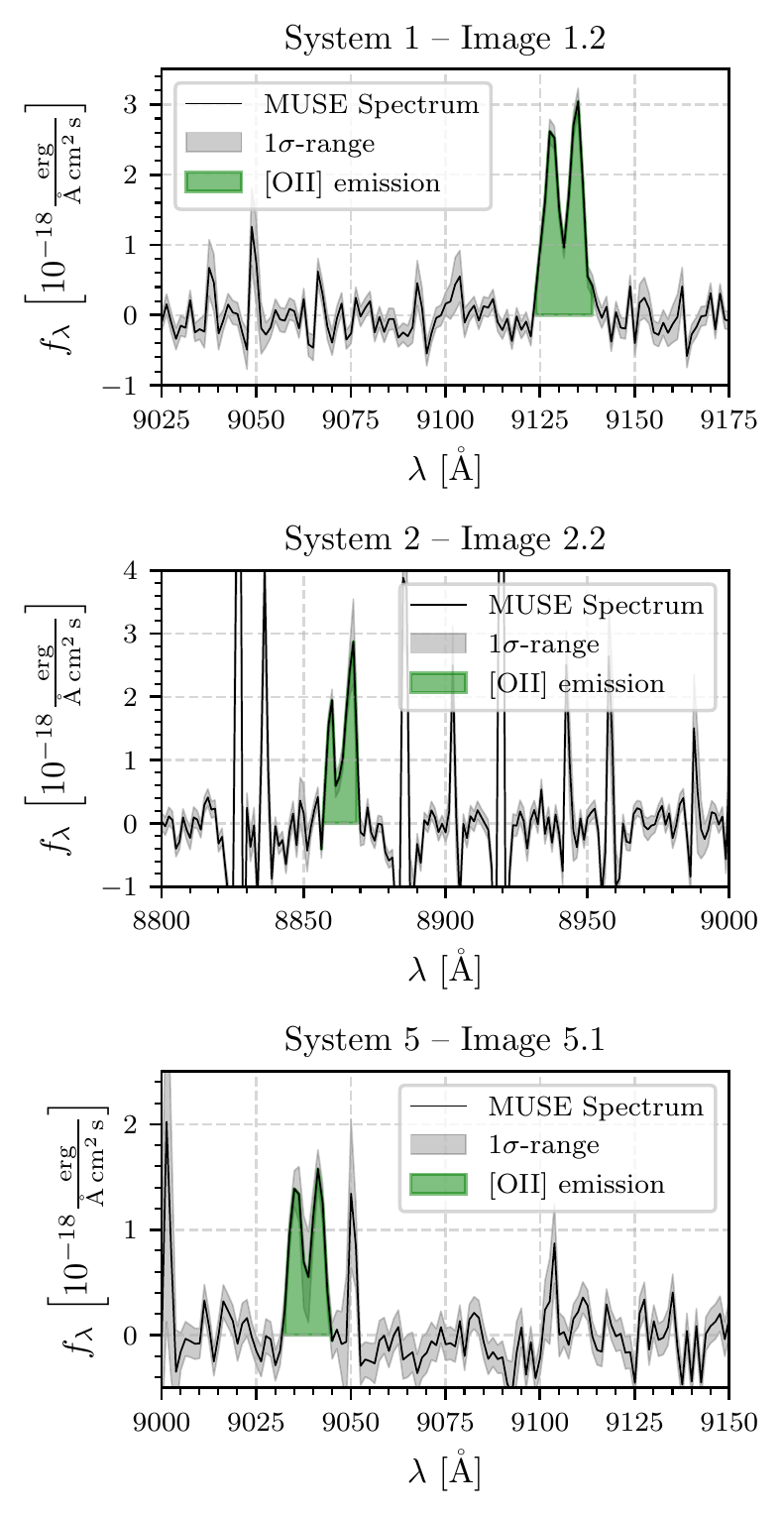}
    \caption{Continuum-subtracted MUSE spectra of one selected image in each of the three spectroscopically confirmed multiple image systems, collapsed to 1D-spectra (black lines). The grey shaded area represents the $1\sigma$-range of the measured spectrum. Each image system shows a strong double-peaked emission feature (green) shaded area that is consistent with the [OII]$\lambda\lambda3726$\AA, $3729$\AA~doublet at redshifts $z\simeq1.45$, $z\simeq1.38$ and $z\simeq1.43$ respectively.}
    \label{fig:MUSE_spectra}
\end{figure}

Before running the lens model minimization, the first task we face is the identification of multiple image families and measuring their redshifts. This is done by first searching for multiple images by eye in the HST data and then identifying the counter images of each system both via their photometric redshift, as measured in the RELICS catalog (see section~\ref{s:data}), and by iterativly using the SL model (see section~\ref{sec:SL-code}) to determine if the images of one system map onto the same source. With this procedure, we identify five multiple image systems in SMACS0723 which are all shown in Fig.~\ref{fig:curve} and Tab.~\ref{multTable}.

Fortunately, we are able to spectroscopically confirm the redshifts of several multiple image systems by scanning the MUSE cube of SMACS0723 (see section~\ref{s:data}) for emission lines. As can be seen in Fig.~\ref{fig:MUSE_images}, systems 1, 2 and 5 show each show a strong double-peaked emission line (see also Fig.~\ref{fig:MUSE_spectra}) consistently over all images in the system in three different wavelength slices. In all three systems, this emission feature is consistent with the [OII]$\lambda\lambda3726$\AA,$3729$\AA~doublet both in terms of peak separation and redshift. Since the doublets are not completely resolved in the MUSE data (see Fig.~\ref{fig:MUSE_spectra}), we fit a simple single Gaussian to the double peaks in order to measure redshifts based on the center between the two peaks. As a result, we measure spectroscopic redshifts of $z=1.450\pm0.001$, $z\simeq1.378\pm0.001$ and $z\simeq1.425\pm0.001$ for systems 1, 2 and 5 respectively. These are in excellent agreement with the photometric redshifts and can also be seen in Tab.~\ref{multTable}. Note that we do not find any significant emission where we would expect the [OII] in the two image candidates 5.3c (see Fig.~\ref{fig:MUSE_images}).

\subsection{Cluster member galaxies} \label{sec:cluster_members}
\begin{figure}
 \begin{center}
  \includegraphics[width=90mm,height=80mm,trim=4.2cm 0.1cm 0.1cm 0.1cm, clip]{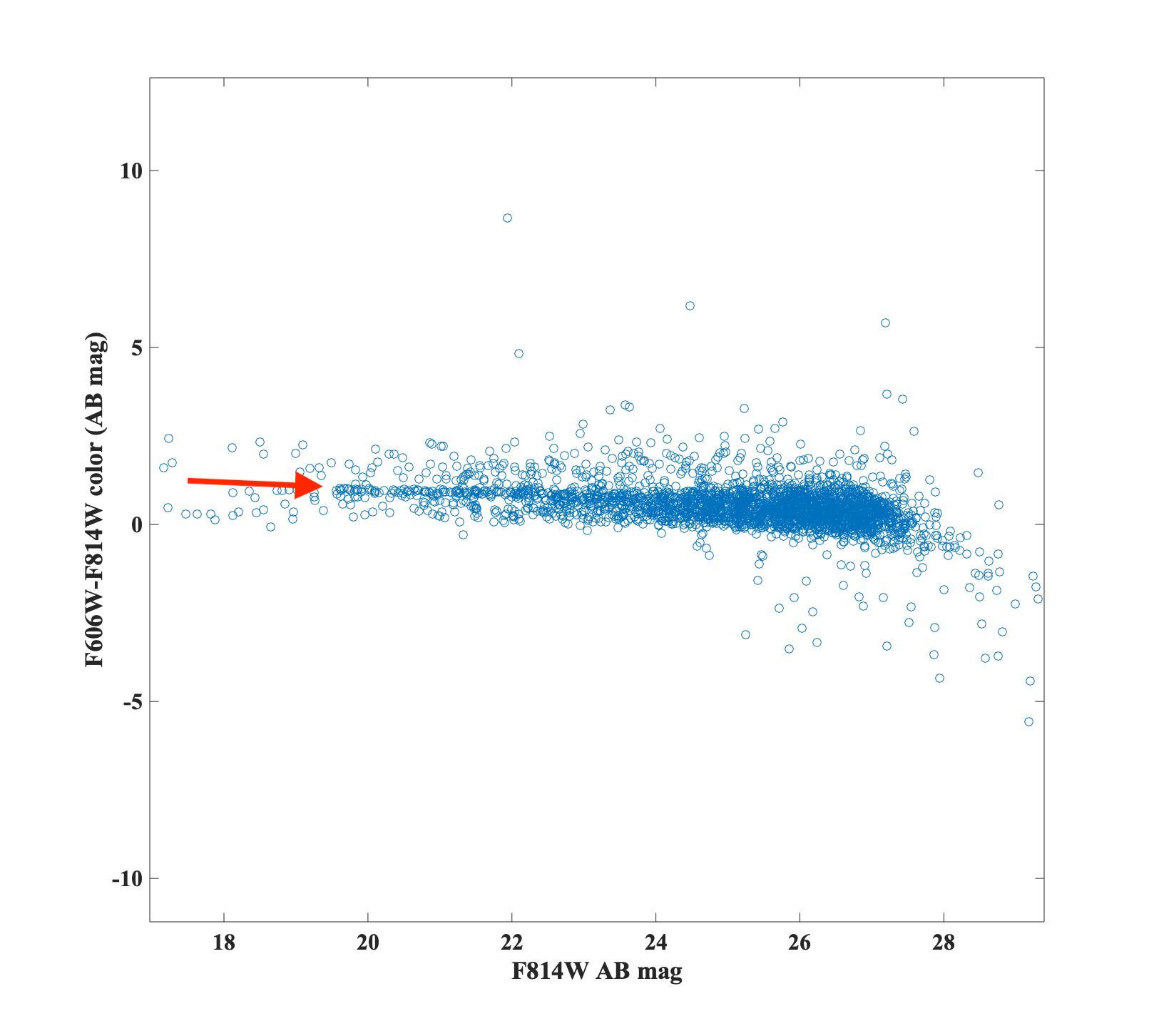}\\
 \end{center}
 \caption{Color-magnitude diagram. The red sequence is clearly visible as a narrow strip. We mark it here with a red arrow for guidance.}
\label{fig:redsequence}
\end{figure}

In order to identify the cluster galaxies for the mass model, we use the red sequence \citep{Gladders2005RCS} of SMACS0723 which stands out in a color-magnitude diagram generated with the F814W and F606W bands (see Fig. \ref{fig:redsequence}). After a by-eye inspection and removal of interloping objects that are not obvious cluster members such as e.g. stars, we remain with 130 galaxies brighter than 23 magnitudes in the F814W band. These will be included as cluster members in our SL model.

\subsection{SL modeling method} \label{sec:SL-code}
We use here the \texttt{Light-Traces-Mass} (\texttt{LTM}) SL modeling method of \citet[][and references therein]{Zitrin2009_cl0024,Zitrin2014CLASH25}, originally based on \citet{Broadhurst2005a}. More complete details of the formalism can be found in the above references and we only give a broad outline here.

A mass model in the LTM formalism we use consist of three main components. The first component of the model is the total mass density distribution of all central cluster galaxies, modeled each with a simple power-law ($q$) surface-mass density scaled by the galaxy's luminosity. The second component is a DM map which is obtained by smoothing the galaxy mass map with a Gaussian kernel of width $S$. The two components are then added with a relative weight $k_{gal}$, reflecting the ratio of luminous to dark matter, and the superposition is scaled to a desired redshift by some factor $K$. The third component, which only contributes to the deflection field but not to the total mass density, is an external shear of strength $\gamma_{ex}$ and position angle $\phi_{ex}$ which allows for greater effective elongation of the critical curves and can help account for the contribution of larger-scale structure. The model thus comprises six main parameters: $q,S,K,k_{gal},\gamma_{ex}$ and $\phi_{ex}$.

Usually we introduce ellipticities and position angles as well as central cores for a few key cluster members, such as the brightest cluster galaxies (which are also the most massive). In addition, it is often useful to leave the relative weight (i.e., the relative mass-to-light ratio) of some key galaxies free as well. Similarly, the redshift of systems lacking spectroscopic measurement, can also be left to be freely optimized.

The optimization of the model is carried out by minimizing a $\chi^2$ function that measures the distance between multiple images and their positions predicted by the model. This is done via a Monte-Carlo Markov Chain (MCMC) with a Metropolis-Hastings algorithm \citep[e.g.,][]{Hastings1970MCMC}. We also include some annealing in the procedure and the chain typically runs for several thousand steps after the burn-in stage. Errors are calculated from the same MCMC chain.

For the modeling of SMACS0723 we leave the relative weight of the 3 brightest galaxies ,including the BCG, to be freely optimized by the model (although for the second and third galaxies we only use a narrow range around the original luminosity). The central core and position angle of the BCG are also left free to be optimized. In addition, the redshifts of systems 3 and 4, whose exact redshifts were not confirmed with MUSE data, are left as free parameters as well around the best-fit photometric redshift. We run a quick model with about 5000 MCMC steps.  

\section{Results and Discussion}\label{s:results}

\begin{figure}
 \begin{center}
  \includegraphics[width=90mm,height=80mm,trim=2.5cm 1cm 1cm 1cm,clip]{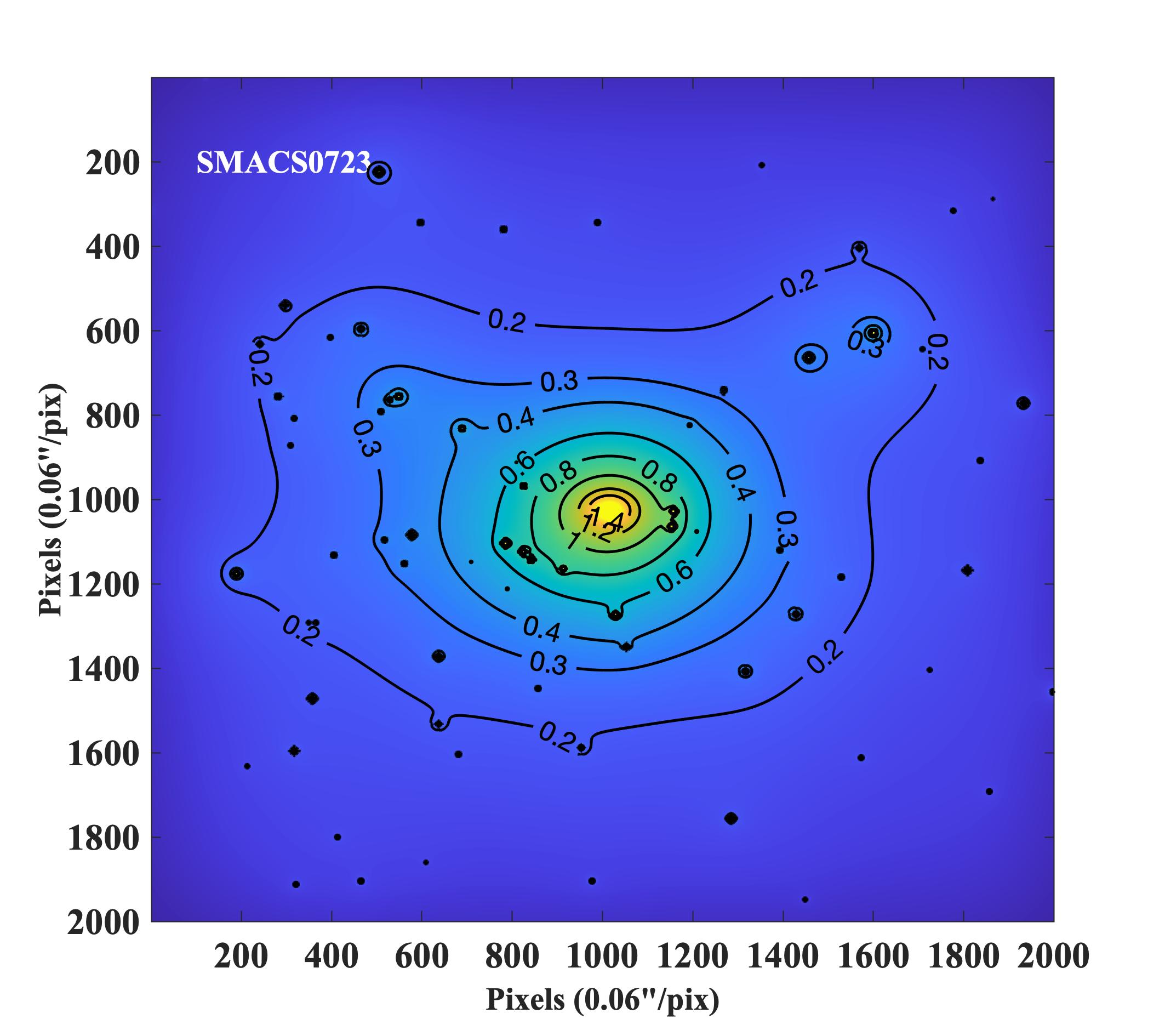}
 \end{center}
 \caption{Projected mass density of SMACS0723. We show
$\kappa$, the surface mass density distribution in units of the critical density for strong lensing, for the source redshift of system 1, at $z=1.45$.}
\label{fig:kappa}
\end{figure}

In Fig. \ref{fig:curve} we show, along with the multiple images, the critical curves of the resulting best-fit \texttt{LTM} model. We note that the model has quite a high \emph{rms} of 2\arcsec.3 in reproducing the position of multiple images. Although the \texttt{LTM} method often gives somewhat higher \emph{rms} values than typical parametric methods (given that it is coupled to the light distribution), the current \emph{rms} is high also compared to the number of systems and simplicity of the lensing system. One of causes may be related to ellipticity: if the underlying projected matter distribution is significantly elongated, the \texttt{LTM} parametrization itself might not be sufficient to easily reproduce it and usually requires a strong external shear, as is indeed the case here (we obtain an extreme external shear value of $\gamma=0.3$). The other reason may be related to the mass-to-light ratio of individual galaxies, which strongly impact the \texttt{LTM} solution. Since individual galaxies deviate in practice from the assumed scaling relation, we often leave the weight of bright galaxies free to be optimized. Here, we essentially only leave the BCG and two other cluster galaxies to be freely weight (within relatively narrow bounds). 
This likely limits the goodness-of-fit of the current model. Future iterations should include a broader range of possibilities, and can be expected to improve the fit.

The projected mass density map, $\kappa$, of the best-fit model is shown in Fig. \ref{fig:kappa}. As can be seen, the mass distribution seems fairly relaxed, with no significant substructure near the center, in line with the relatively smooth and symmetric critical curve. Note however, that given that the ellipticity of the \texttt{LTM} is limited, the critical curves in practice may be more elongated or more elliptical. Thus, in future works it would be interesting to compare to parametric models explicitly, in which the ellipticity can be assigned intrinsically in the mass distribution and is thus not as severely limited. In other words, we expect that analytic models would show a more elongated curve than is seen in Fig. \ref{fig:curve} here.

One other feature of lensing clusters that is useful to assess the lensing efficiency is the Einstein radius. The effective Einstein radius we find is modest, $14.5\pm2\arcsec$ for a source at $z_{s}=1.45$, where the effective Einstein radius is defined as the radius of the area enclosed within the critical curves if it were a circle. The mass in that critical area is $3.42\pm0.47\times10^{14}$ M$_{\odot}$. For source redshift of $z_{s}=2$ we find $\theta_{E}=16.9\pm2\arcsec$, enclosing $4.15\pm0.58\times10^{14}$ M$_{\odot}$. Errors on the Einstein radius and mass are nominal, systematic values, reflecting typical errors seen between different models; the statistical uncertainties are somewhat smaller. These estimates will be revised when more multiple images are identified in JWST data.

\begin{figure*}
 \begin{center}
  \includegraphics[width=0.45\textwidth,trim=0cm 0cm 0cm 0cm,clip]{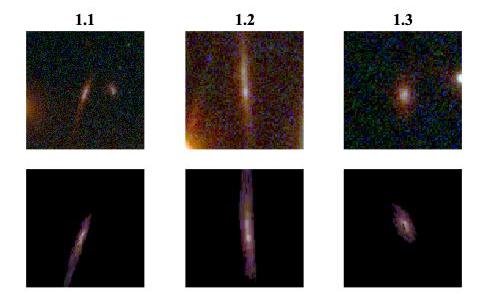}
  \includegraphics[width=0.43\textwidth,trim=0cm 0cm 0cm 0cm,clip]{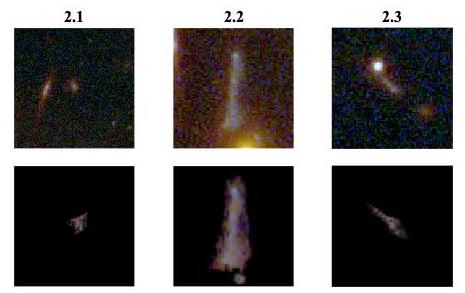}\\
   \includegraphics[width=0.5\textwidth,trim=0cm 0cm 0cm 0cm,clip]{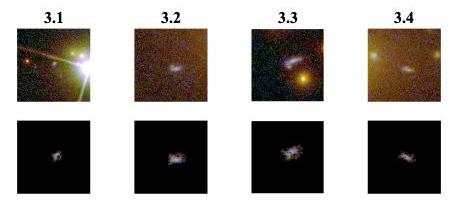}\\
   \includegraphics[width=0.4\textwidth,trim=0cm 0cm 0cm 0cm,clip]{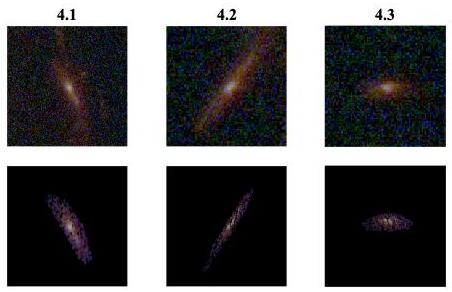}
   \includegraphics[width=0.42\textwidth,trim=0cm 0cm 0cm 0cm,clip]{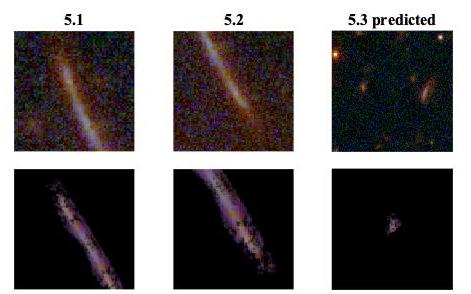}
 \end{center}
\caption{Reproduction of multiple images by our model. For each system we lens one of the arcs to the source-plane and back through the lens to reproduce the other images of the systems. Note images are slightly smoothed and centered for better view (and thus their location is not an indication to the accuracy of the model). The upper row for each system shows stamp cutouts of the original images and the bottom row the reproduction by the model. In system 5, the third image is predicted but two options are seen nearby. Overall, the model seems to reproduce very well the multiple images.}\vspace{0.1cm}
\label{fig:Reporudiction}
\end{figure*}

\section{Summary}\label{s:summary}
We presented a new HST SL model for the massive cluster SMACS0723, that -- according to media -- was recently imaged with JWST and whose data are to be published in the next few days. 

We find a relatively modest lens of $\theta_{E}\simeq 14.5 \pm 2\arcsec$ for a source at $z=1.45$ and $\theta_{E}\simeq 16.9 \pm 2\arcsec$ for a fiducial source at $z=2$. The mass distribution seems to be relatively relaxed with no major dark matter sub-structure close to the cluster center. 

The importance of our lens model is two-fold: First, it is constructed with a distinct method which is different than typical parametric modeling techniques, and thus probes a different part of the solution space. This will become useful for upcoming studies of the high-redshift galaxy population behind SMACS0723 using JWST data. Second, it is made public right before the release of the JWST data of this cluster. Because of the greater depth of JWST observations, the released imaging data might reveal new multiple image systems (and high-redshift galaxies) which in turn would increase the number of constraints on the DM distribution of the cluster. Given that our current model is ``JWST-blind", it will allow for a comparison with future versions that will make use of the new JWST data.

\section*{acknowledgements}
We wish to thank the RELICS, BUFFALO, ALCS, and PEARLS collaborations for data products or discussions that have stimulated and enabled this work. We also thank Johan Richard and Hakim Atek for very useful discussions.

The BGU group acknowledges support by Grant No. 2020750 from the United States-Israel Binational Science Foundation (BSF) and Grant No. 2109066 from the United States National Science Foundation (NSF), and by the Ministry of Science \& Technology, Israel. AZ acknowledges useful comments and support by Ely Kovetz. 

This work is based on observations made with the NASA/ESA Hubble Space Telescope obtained from the Space Telescope Science Institute, which is operated by the Association of Universities for Research in Astronomy, Inc., under NASA contract NAS 5–26555. These observations are associated with program ID 15959. Support for program 15959 was provided by NASA through a grant from the Space Telescope Science Institute, which is operated by the Association of Universities for Research in Astronomy, Inc., under NASA contract NAS 5-26555. This work is also based on observations made with ESO Telescopes at the La Silla Paranal Observatory obtained from the ESO Science Archive Facility.

This research made use of \texttt{Astropy},\footnote{\url{http://www.astropy.org}} a community-developed core Python package for Astronomy \citep{astropy13,astropy18} as well as the packages \texttt{NumPy} \citep{vanderwalt11}, \texttt{SciPy} \citep{virtanen20}, \texttt{matplotlib} 
\citep{hunter07}, \texttt{specutils} \citep[][]{specutils21}, \texttt{spectral-cube} \citep{spectral-cube14} and some of the astronomy \texttt{MATLAB} packages \citep{maat14}.


\end{document}